# Permeability changes in Boom clay with temperature

P. Delage[1], N. Sultan[2], Y.J. Cui[1], Li X.L.[3]

[1] Ecole des ponts ParisTech, UR Navier/CERMES,
6-8 av. B. Pascal, F 77455 Marne la Vallée cdx 2, France
[2] IFREMER, BP 70, 28280 Plouzané, France
[3] EURIDICE Group, c/o SCK-CEN, Mol Belgium

**Summary**

In the framework of research into radioactive waste disposal, this paper presents some tests carried out to investigate the effects of temperature on the changes in permeability of Boom between 20 and 90°C. Constant head permeability tests were carried out in a high pressure isotropic compression tests at various temperature and isotropic stresses. The results show that the changes in permeability of Boom clay with temperature are only due to the changes in viscosity of free water with temperaure. This demonstrates on the one hand that the water involved in transfer at various temperatures is free water and on the other hand that there is apparently only few changes in the status of adsorbed water with respect to temperature between 20 and 90°C.

## 1. Introduction

Relatively few data are available about the changes in permeability in clays with temperature. This problem is of significant importance when considering the feasibility of high level radioactive waste disposal in clays at great depth. It is well known that the changes in permeability in inactive porous media are directly related to the changes in viscosity of water with temperature. This is less evident in clay and particularly plastic clays, in which significant clay-water interaction exists. It is known that clay-water interactions are sensitive to temperature changes and the temperature dependence of permeability in these clays hence requires further investigation.

Permeability tests at various temperatures were carried on Boom clay samples extracted from the HADES facility in Mol (Belgium). Constant head tests were carried out at various temperatures between 20 and 100°C. by applying a constant pressure gradient across a standard triaxial sample by using the pressure-volume controllers. Tests were carried out in a high pressure temperature controlled isotropic cell in which confining and back pressures were controlled by using digital pressure-volume controllers.

## 2. Materials and methods

Tests were performed on samples, extracted from a depth of 223 m in the Boom clay deposit (Decleer et al. 1983), in the underground facilities of SCK-CEN in Mol (Belgium). Boom clay is a stiff clay, with a plasticity index of about 50%, a natural porosity around 40% and a water content between 24 and 30%. Previous work on the thermal behaviour of Boom clay includes Baldi et al. (1988), De Bruyn and Thimus (1995), Belanteur et al. (1997), Delage et al. (2000), Cui et al. (2000), Sultan et al. 2002, Cui et al. (2009).

Tests were carried out in an isotropic compression cell (see Figure 1) designed to support high pressures and high temperatures (up to 60 MPa and 100°C respectively). A heating coil was placed on the outer wall of the cell, and the temperature was controlled by a thermocouple thermometer placed inside the cell, in the confining water. The precision of the temperature regulation system was ±0.05°C. The isotropic stress was applied by a high pressure volume controller (PVC, GDS brand, 60 MPa) and the back pressure (up to 2 MPa) by a standard PVC. The main advantage of this system is its ability to monitor volume changes while applying pressures. Due to the thermal volume changes of water under non isothermal conditions, it was decided to monitor the sample volume changes from the changes in volume of the confining water. As discussed in Delage et al. (2000), this technique is much more satisfactory at high stresses than in the range of stresses used in standard soil mechanics.

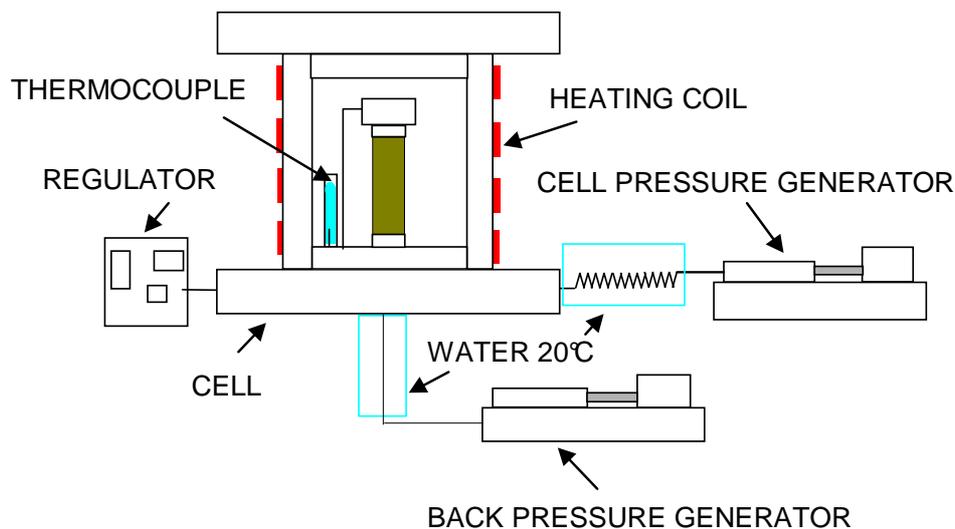

Figure 1. Schematic diagram of the triaxial cell.

## 3. Experimental results

In order to ensure significant flow rates, a sample smaller than the standard triaxial sample was used. The standard diameter ($\phi = 37.7$ mm) was kept but the height was reduced down to $h = 22.95$ mm. The sample was saturated under a low all round pressure (110 kPa) with a back pressure equal to 40 kPa. A Skemton $B$ coefficient of 0.97 was attained after 24 hours. Afterwards, the sample was consolidated up to 2.5 MPa.

The first permeability measurement was carried out at T = 20°C by increasing the back pressure up to 1 MPa at the bottom of the sample, while putting the top porous stone in contact with atmospheric pressure. It results in a high value of the hydraulic gradient ($i \approx 4000$), as compared to that of variable head permeameters ($i \approx 100$). The high gradient was found necessary in order to create a measurable flow in the dense plastic clay, and to get a satisfactory precision in the measurement of flow rate and, hence, of the permeability.

In Figure 2, the volume of water injected at the base is given as a function of time. In spite of the high gradient applied, the figure shows that 10 hours were necessary to achieve permanent flow and 15 hours to obtain a satisfactory determination of the slope of the curve. This section corresponds to a constant flow from which the permeability can be calculated. At 20°C, a permeability value of 2.5 $10^{-12}$ m/s was obtained with a porosity value equal to 39%. Possible problems related to the high value of the hydraulic gradient were examined by comparing this result with that of a variable head test, run in a rigid oedometer cell ($i \approx 100$). A permeability value of 3.5 $10^{-12}$ m/s for a porosity of 44.1% was obtained, a result consistent with the former result. The two points are plotted in Figure 3.

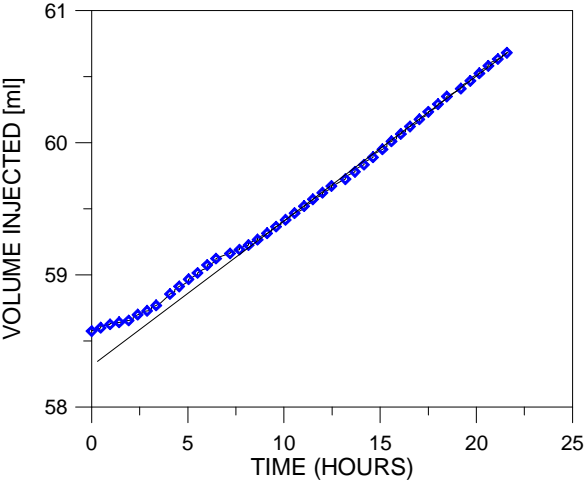

Figure 2. Volume of water injected during a permeability test

The test carried out for the determination of the permeability at various temperatures is presented in Figure 3, in a diagram giving the porosity versus the logarithm of the permeability. Constant head permeability tests were successively performed under an all-round pressure of 2.5 MPa at 20°C, 60°C, 70°C, 80°C and 90°C. Heating phases were very progressive, at a heating rate of 0.1°C/20 minutes, in order to fulfil drained conditions (Cui et al. 2000). At 90°C, the sample was isothermally loaded from 2.5 up to 4 MPa, and permeability tests were carried out in a cooling phase at 90°C, 80°C, 70°C, 60°C and 30°C. At 30°C, subsequent isothermal loading from 4 MPa up to 6 MPa was carried out, and permeability tests were performed at 30°C and 60°C.

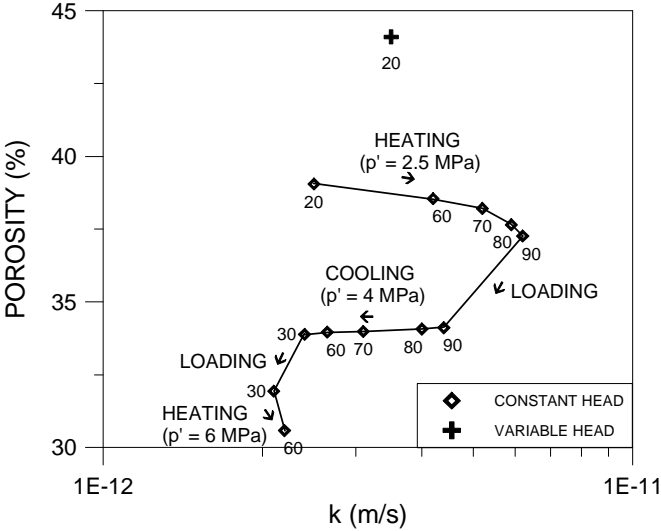

Figure 3. Permeability tests performed on a sample of Boom clay at various temperatures and stresses.

One observes in Figure 3 that heating from 20°C up to 90°C results in a contraction of the sample (the porosity $n$ decreasing from 39 % down to 37.2 %) and in an increase in permeability from $2.5 \, 10^{-12}$ m/s to $6.2 \, 10^{-12}$ m/s. The loading sequence at 4 MPa at a temperature of 90°C decreases the permeability down to $4.4 \, 10^{-12}$ m/s. During the cooling phase under 4 MPa, the permeability decreases from $4.4 \, 10^{-12}$ m/s down to $2.4 \, 10^{-12}$ m/s and the porosity remains almost constant (34 % to 33.9 %). Loading up to 6 MPa under a temperature of 30°C decreases the porosity down to 31.9%. Further heating at 60°C under 6 MPa gives a final porosity of 30.6%.

The changes in permeability presented in Figure 3 are due to the coupled effect of changes in temperature and porosity. In order to separate these effects, the intrinsic permeability values ($K$) were calculated according to :

$$k = \frac{K \gamma_w(T)}{\mu(T)} \qquad (1)$$

where $\mu(T)$ is the water viscosity and $\gamma_w(T)$ is the unit weight of water. The following relation, valid for free water, was derived from experimental values reported by Hillel (1980):

$$\mu(T) = -0.00046575 \ln(T) + 0.00239138 \, (Pa.s) \qquad (2)$$

In the range of temperatures considered, there is no significant change in the value of $\gamma_w(T)$.

Observation of the results presented in Figure 4 in a semi-logarithmic diagram shows that the relationship between the intrinsic permeability $K$ and the porosity $n$ is unique and linear. It appears reasonably independent of temperature. The data obtained from the variable head test performed in the rigid ring oedometer also correctly matches with the results. Consequently, the intrinsic permeability of a sample loaded at a given temperature is only dependent on its porosity, independently on the *thermo*-mechanical path previously followed in a $(p';T)$ plane. In other words, volume changes created by stress and/or temperature changes have the same effect on the intrinsic permeability of a sample. Data of Figure 4 are also consistent with the interpretation of Habibagahi (1977) and confirm that the water put in movement during a permeability test is the free water that circulates through the channels limited by the water tightly adsorbed on the clay minerals.

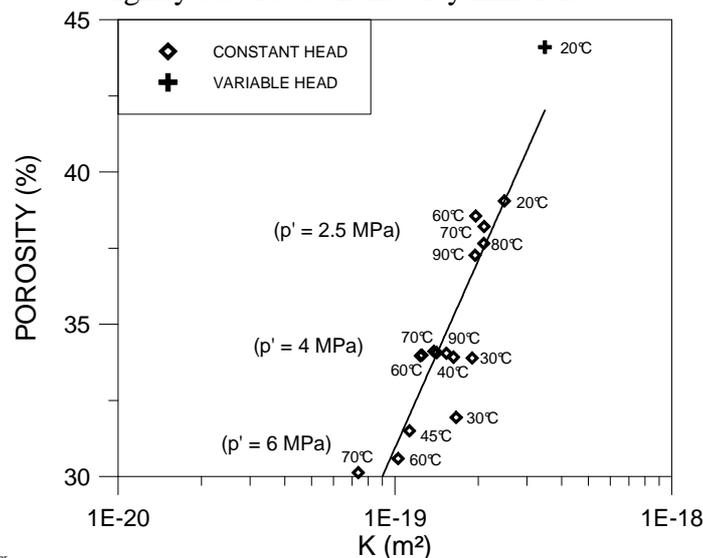

Figure 4. Results of the permeability tests, in terms of intrinsic permeability.

Similar conclusions regarding the intrinsic permeability were drawn by Morin and Silva (1984) from permeability tests carried out at various temperatures (between 22 and 220°C) on 4 soils of different plasticity indices and higher void ratios ($I_p$ between 52 and 179, and $e$ between 1 and 9). They also showed that, at a given void ratio, a more plastic soil is more impervious, due to a greater number of layers of water molecules adsorbed on the clay minerals, which do not participate to the flow, as commented by Habibagahi (1977).

## 4. Conclusion

Constant head permeability tests carried out on Boom clay samples at various temperatures between 20° and 90°C and under various isotropic stresses demonstrated that the permeability changes observed were only linked to changes in the viscosity of free water with temperature. As a consequence, no temperature influence was observed on the changes in intrinsic permeability with temperature. This means firstly that the water molecules involved in the transfers at various temperature are free molecules and secondly that there is no significant changes in the status of adsorbed water with temperature between 20 and 90°C.